# Foundations of Non-linear Quantum Mechanics

Edward Rówiński[1](*)

[1]Institute of Materials Science, University of Silesia, 75 Pułku Piechoty Street 1A, 41-500 Chorzów, Poland

(*) edward.rowinski@us.edu.pl

**Abstract**

The foundations of non-linear quantum mechanics are based on six postulates and five propositions. On a first quantised level, these approaches are built on non-linear differential operators, non-linear eigenvalue equations, and the notion of non-linear observables and non-linear states. The present theory predicts that the non-linear function solution of a non-linear partial differential equation for a free particle is correct and that the commutator of the position operator and the non-linear momentum operator does not commute. Moreover, the implicit wave function of linear quantum mechanics can be determined by a non-linear wave function solution of non-linear partial differential equation, which also verifies the non-linear quantum mechanics.

**Introduction**

Weinberg presents a general framework for introducing non-linear corrections into ordinary quantum mechanics that can serve as a guide to experiments sensitive to such corrections [1]. In the class of generalised theories described here, the equations that determine the time-dependence of the wave function are no longer linear, but of the Hamiltonian type. Also, wave functions that differ by a constant factor represent the same physical state and satisfy the same time-dependence equations. As a result, there is no difficulty in combining separate sub-systems. Prescriptions are given for determining the states in which observables have definite values, and for calculating the expectation values of observables for general states, but the calculation of probabilities requires detailed analysis of the method of measurement. A study is presented of various experimental possibilities, including the precession of spinning particles in external fields, experiments of Stern-Gerlach type, and the broadening and de-tuning of absorption lines.

It is proposed to set stringent limits on possible non-linear corrections to ordinary quantum mechanics by searching for the de-tuning of resonant transitions [2]. A suggested non-linear generalisation of quantum mechanics is used to show that such de-tuning would be expected in the rf transition in $^9Be^+$ ions that is used to set frequency standards. Measurements at the National Bureau of Standards already set limits of order $10^{-21}$ ion the fraction of the energy of the $^9Be$ nucleus that could be due to non-linear corrections to quantum mechanics, with good prospects of improving this by two to three orders of magnitude.

The Gross-Pitaevski equation, or non-linear Schrodinger equation, are widely used in the theory of superfluidity and in the theory of Bose-Einstein condensation [3]. In quantum field theory, non-linearity occurs in the equations of interacting field operators [4]. Here, the field operators remain linear. Non-linear quantum mechanics does not have a large literature and its content is very varied [3–20].

In applications of quantum mechanics, where the several operators involved act in space and time variables with derivatives of different orders, solving the problem of non-linearity requires more than the modified Schrodinger equation or the Weinberg non-linear corrections. Using the same physical tools, many of the results familiar in the theory of linear operators could be extended to non-linear operators.

The paper presents further progress in this area and focuses on the foundations of non-linear quantum mechanics (NQM) based on the six major postulates and five propositions.

**Results and Discussion**

One of the motivations is based on the view that any foundations of non-linear quantum mechanics in a non-relativistic approach must contain the non-linearity arising from the standard wave function. Here, a remark is necessary on a composition of two functions, when one function is inside of another function. In order to derive postulates and propositions of non-linear quantum mechanics, we may start with the composition of two functions. For this reason, the modified definition of a non-linear function is given by:

**Definition 1.** *The composition of two functions is defined by*

$$(\psi_{non,x} \circ \phi_x)(x) := \psi_{non,x}[x, \phi_x(x)] \qquad (1)$$
$$(\psi_{non,t} \circ \phi_t)(t) := \psi_{non,t}[t, \phi_t(t)] \qquad (2)$$

*and*

$$\psi_{non,t}[t, \phi_t(t)] \cdot \psi_{non,x}[x, \phi_x(x)] := \psi_{non}[t, x, \phi_{x,t}(t,x)] \equiv \psi_{non}(t, x, \phi_{x,t}) \qquad (3)$$

*where $(\psi_{non,x} \circ \phi_x)(x)$ – the composition of the function $\psi_{non,x}$ with $\phi_x$, $\psi_{non,x}$ – the position part of the non-linear function, $\phi_x (= \phi_x(x))$ – the time-independent wave function, $x$ – the Cartesian coordinate, $(\psi_{non,t} \circ \phi_t)(t)$ - the composition of the function $\psi_{non,t}$ with $\phi_t$, $\psi_{non,t}$ – the time part of the non-linear function, $\phi_t (= \phi_t(t))$ – the time-dependent wave function, $t$ – the time, $\psi_{non}[t, x, \phi_{x,t}(t,x)]$ - the non-linear function in $\phi_{x,t}(t,x)$, $\phi_{x,t} (= \phi_{x,t}(t,x))$ – the wave function.*

This definition is a specific physical approach, which leads to the general concept of the non-linearity of the form $\psi_{non} = \psi_{non}(t, x, \phi_{x,t})$. It is well known that the Hilbert space combines the properties of two fundamentally different types of spaces: a vector space and a metric space [21]. Note that in the case of the vector space, the ket vector is given by $|\psi_{non,x}>$, where $|\psi_{non,x}>$ - the ket vector of Dirac's notation. When this notation is used, the wave function of $<x|\phi_x>$ splits into the bra ($<x|$) and the ket ($|\phi_x>$) vectors. Thus the outer product of $|x> \otimes |\phi_x>$ is $|x, \phi_x>$, where $|x, \phi_x>$ - the ket-vector notation for tensor space. The Hermitian conjugate of the ket vector is a bra vector: $|x, \phi_x>^+ = <x, \phi_x|$. The $<x, \phi_x|$ is defined by its action on $|\psi_{non,x}>$ as follows $<x, \phi_x|: |\psi_{non,x}> \to <x, \phi_x|\psi_{non,x}> \equiv \psi_{non,x}(x, \phi_x)$. Thus we obtain the desired result. In the case of the metric space the distance $D(\psi_{1,non}, \psi_{2,non})$ between any two non-linear functions ($\psi_{1,non}, \psi_{2,non}$) is defined as

$D(\psi_{1,non}, \psi_{2,non}) = \int \sqrt{|\psi_{1,non}|^2 + |\psi_{2,non}|^2 - 2\psi_{1,non}\psi_{2,non}} \, dx$. The distance between any two N-particle non-linear functions is metric. The space of all non-linear functions, with the metric, becomes a metric space. The major point here is that there is more than one language in which to express nonlinear quantum mechanics. The most common language for the non-linear theory is the state function, but momentum space offers an equally important view of the state function. The composition of the non-linear function with the wave function can be regarded as the correspondence principle between the non-linear and linear quantum mechanics. Moreover, the implicit wave function can be determined by the non-linear function solution of non-linear partial differential equation, which also verifies non-linear quantum mechanics (Postulate 4, Proposition 5.1). Let us start with an enumeration of the basic postulates of NQM. The postulates and propositions read as follows:

**Postulate 1.** (Non-linear wave function). *The non-linear function of $\psi_{non} = \psi_{non}(t, x, \phi_{x,t})$ is called the non-linear wave function if the probability of finding a single particle* somewhere *is 1, so that the normalisation condition is*

$$\int_{-\infty}^{\infty} \psi^*_{non}(t, x, \phi^*_{x,t}) \psi_{non}(t, x, \phi_{x,t}) dx = 1 \qquad \text{for all times} \qquad (4)$$

*where $\psi^*_{non}(t, x, \phi^*_{x,t})$ – the complex conjugate of $\psi_{non}(t, x, \phi_{x,t})$. It is customary to also normalise many-particle functions to 1.*

Let $\psi^*_{non}(x, \phi^*_x) = C_x \exp[\exp(-ix/\lambdabar)]$ and $\psi_{non}(x, \phi_x) = C_x \exp[\exp(ix/\lambdabar)]$ be the non-linear complex functions, where $\lambdabar = \lambda_B/(2\pi)$, $\lambda_B$ – the de Broglie wavelength of the particle, $C_x$ – the constant. The normalisation condition is

$$|C_x|^2 \int_{-\infty}^{\infty} \delta[g(x)] dx = 1 \qquad (5)$$

where $\delta[g(x)]$ – the delta function, $g(x)$ – the real function. If the Maclaurin series expansion for $g(x) = \exp[2\cos(x/\lambdabar)]$ is the first three terms, then $\delta[g(x)] = 1/(2\lambdabar)\delta(x - \lambdabar)$, for $\lambdabar \geq 0$, where $\lambdabar$ – the positive root of $x^2 - \lambdabar^2 = 0$, $g'(x)$ – the derivative of $g(x)$ with respect to $x$ at the fixed $\lambdabar$. Finally, this gives

$$\frac{|C_x|^2}{2\lambdabar} \int_{-\infty}^{\infty} \delta(x - \lambdabar) dx = 1 \Rightarrow \frac{|C_x|^2}{2\lambdabar} = 1 \Rightarrow C_x = \sqrt{2\lambdabar}. \qquad (6)$$

So we obtain the desired result. Moreover, the second postulate is due to the non-linear differential operators, where a quantum differential operator acts on the non-linear function.

**Postulate 2**. (Non-linear differential operator). *Let $\psi_{non} = \psi_{non,x}(x, \phi_x)$ ($\psi_{non}(t, \phi_t)$) be a non-linear function in $\phi_x$ ($\phi_t$), where $\phi_x = \phi_x(x)$ ($\phi_t = \phi_t(t)$) is the eigenfunction of $\hat{p}_x$ ($\hat{E}$). Then, the quantum differential operators which act on the non-linear function are the non-linear differential operators.*

The momentum operator in the position representation, $\hat{P}_{(x),non} = (\hbar/i)(\partial/\partial x)$, is said to be non-linear if

$$\hat{P}_{(x),non}(\alpha \cdot \psi_{non,x} + \beta \cdot \varphi_{non,x}) = \frac{\hbar}{i}\frac{\partial}{\partial x}(\alpha \cdot \psi_{non,x} + \beta \cdot \varphi_{non,x}) =$$

$$= \alpha \frac{\hbar}{i}\frac{\partial \psi_{non,x}}{\partial \phi_x}\frac{\partial \phi_x}{\partial x} + \beta \cdot \frac{\hbar}{i}\frac{\partial \varphi_{non,x}}{\partial \phi_x}\frac{\partial \phi_x}{\partial x} = \alpha \frac{\partial \psi_{non,x}}{\partial \phi_x}\frac{\hbar}{i}\frac{\partial \phi_x}{\partial x} + \beta \cdot \frac{\partial \varphi_{non,x}}{\partial \phi_x}\frac{\hbar}{i}\frac{\partial \phi_x}{\partial x} = \quad (7)$$

$$= \alpha \cdot \frac{\partial \psi_{non,x}}{\partial \phi_x}\cdot \hat{p}_x\phi_x + \beta \cdot \frac{\partial \varphi_{non,x}}{\partial \phi_x}\cdot \hat{p}_x\phi_x$$

where $\alpha$ and $\beta$ – the constants, $\varphi_{non,x}$ – the non-linear function, $\partial \psi_{non,x}/\partial \phi_x$ and $\partial \varphi_{non,x}/\partial \phi_x$ – the non-linear factors, $\hat{p}_x$ – the Hermitian operator of momentum. The non-linear momentum operator $\hat{P}_{(x),non}$ is constructed as the product of two operators: the unique operator of $\partial/\partial \phi_x$ and the momentum operator of $\hat{p}_x$. The non-linearity is due to the non-linear factors. The third fundamental postulate will introduce the way of setting up an eigenvalue equation, e.g. for momentum and energy.

**Postulate 3**. (Non-linear eigenvalue equation). *Given a non-linear transformation $\hat{P}_{(\cdot),non} = -i\hbar \partial/\partial x; \quad i\hbar \partial/\partial t; \quad ...$, a function*

$$\psi_{non} = \begin{cases} C_x \cdot \exp(\phi_x) = C_x \cdot \exp[\exp(\frac{i}{\hbar}p_x x)] \\ C_t \cdot \exp(\phi_t) = C_t \cdot \exp[\exp(\frac{i}{\hbar}Et)] \end{cases} \quad (8)$$

*is defined to be a non-linearity of the transformation if it satisfies the non-linear eigenvalue equations*

$$\hat{P}_{(x),non}\psi_{non} = p_x\psi_{non}\phi_x; \quad (9)$$
$$\hat{P}_{(t),non}\psi_{non} = E\psi_{non}\phi_t; \quad (10)$$
...

*with*
$$\hat{P}_{(\cdot),non}\psi_{non} \equiv \hat{A}\psi_{non}\hat{B}_Q\phi_x \quad (11)$$
$$\hat{B}_Q = \hat{p}_x, \hat{E},..., \quad \hat{A} = \frac{\partial}{\partial \phi_x} \quad (12)$$

*and the linear eigenvalue equations*

$$\hat{p}_x\phi_x \equiv \frac{\hbar}{i}\frac{\partial \phi_x}{\partial x} = p_x\phi_x, \quad \hat{E}\phi_t \equiv -\frac{\hbar}{i}\frac{\partial \phi_t}{\partial t} = E\phi_t, \quad ... \quad (13)$$

*for some scalar $\lambda = p_x, E, ...$, where $\hat{P}_{(x),non}$ ($\hat{P}_{(t),non}$) – the non-linear operator of momentum (energy), $\hat{P}_{(\cdot),non} = \hat{P}_{(x),non}, \hat{P}_{(t),non},..., p_x$ ( $E$ ) – the eigenvalue of $\hat{p}_x$ ( $\hat{E}$ ), $\hat{A}$ – the unique differential operator acting on the non-linear function, $\hat{B}_Q$ – the quantum differential*

*operator acting on the wave function, $\hat{E}$ – the Hermitian operator of energy, $\hbar$ – the reduced Planck's constant, $C_x$, $C_t$ – the constants.*

From the derived equations, it follows that the unique differential operator ($\hat{A}$) maps the non-linear function $\psi_{non}$ to the constant function. The non-linear operator $\hat{P}_{(x),non}$ is constructed as the product of two operators: the unique operator of $\partial/\partial\phi_x$ and the momentum operator of $\hat{p}_x$. Mathematically, let $X$ and $Y$ be two vector spaces and a mapping $\hat{P}_{(x),non}: X \to Y$ over a common field of scalars that does not have the property of linearity. If $Y$ is the set $\Re$ of real or $\Im$ of complex numbers, then a non-linear operator is called a non-linear functional. The non-linear momentum eigenstate is

$$\psi_{non}(x,\phi_x) = \sqrt{\hbar/p_x} \cdot \exp(\phi_x) = \sqrt{\hbar/p_x} \cdot \exp[\exp(\frac{i}{\hbar} p_x x)] \tag{14}$$

with the normalisation condition

$$|C_x|^2 \int_{-\infty}^{\infty} \delta\{\exp[2\cos(\frac{p_x x}{\hbar})]\}dx \cong \frac{|C_x|^2 p_x}{\hbar} \int_{-\infty}^{\infty} \delta(x - \hbar/p_x)dx = 1 \Rightarrow C_x = \sqrt{\hbar/p_x}. \tag{15}$$

Finally, the non-linear eigenvalue equation of momentum is approximated by

$$\frac{\hbar}{i} \frac{\partial \psi_{non}}{\partial x} = p_x \left\{ \sqrt{\hbar/p_x} \cdot \exp\left[\frac{i}{\hbar} p_x x + \exp(\frac{i}{\hbar} p_x x)\right] \right\}. \tag{16}$$

The eigenstates of $\hat{P}_{(x),non}$ are the product form of the momentum eigenfunction and the non-linear eigenfunction with the eigenvalue of $p_x$. The expectation value of the observable $P_{(x),non}(\psi_{non})$ is given by

$$P_{(x),non}(\psi_{non}) := \int_{-\infty}^{\infty} \psi_{non}^* \hat{P}_{(x),non} \psi_{non} dx = p_x \tag{17}$$

The expectation value of the non-linear observable $P_{(x),non}(\psi_{non})$ predicts the same result as in the linear quantum mechanics. The first alternative of Kibble for a non-linear extension of standard quantum theory is to replace the observable $O(x)$ by a general function on the tangent space of $\Gamma$ [19].

**Proposition 3.1.** (Expectation value of NQM). *The expectation value of an observable $O(\psi_{non})$ is given by*

$$O(\psi_{non}) := <\psi_{non}|\hat{O}|\psi_{non}> \tag{18}$$

*where $|\psi_{non}>$, $<\psi_{non}|$ – the ket and bra vectors in Dirac's notation, $\hat{O}$ – the operator.*

**Proposition 3.2**. *The probability of obtaining a non-linear momentum eigenvalue at fixed point* $(|\hat{A}\psi_{non}| = const)$ $a_j = const \cdot p_x$, *in a measurement of an observable* $P_{(x),non}$ *is given by the square of the inner product of* $|\psi_{non}>|\phi_x>$ *with the state* $|a_j>$, $|<a_j \| \psi_{non}>|\phi_x>|^2$.

**Proposition 3.3**. *Immediately after the measurement of an observable* $P_{(x),non}$ *has yielded a value* $a_j = const \cdot p_x$, *the state of the system is the normalised eigenstate* $|a_j>$.

The states are to be normalised to unity: $<\phi_x|\phi_x> = 1$ and $<a_j|a_k> = \delta_{ij}$. Sometimes, this is not possible. When acting with the squared position operators of $\hat{x}^2$ on $\psi_{non}$ we obtain $\hat{x}^2 \psi_{non} = x^2 \psi_{non}$. They have the same forms as the operators discussed in the linear quantum mechanics. Since quantum mechanics is so useful in the case of linear operators, it is natural to try to extend its principles to non-linear operators. A new differential equation for non-linear functions (or non-linear wave functions) is the form of the non-linear eigenvalue equation, which is given by postulate 4.

**Postulate 4**. *The time evolution of a non-linear function is given by a continuous non-linear partial differential equation (NPDE) for a single particle of mass m:*

$$i\hbar \frac{\partial \psi_{non}}{\partial t} = -\frac{\hbar^2}{2m}\frac{\partial^2 \psi_{non}}{\partial x^2} + \hat{V}\psi_{non} \tag{19}$$

with

$$i\hbar \frac{\partial \psi_{non}}{\partial t} = i\hbar \frac{\partial \psi_{non}}{\partial \phi_{x,t}}\frac{\partial \phi_{x,t}}{\partial t} \tag{20}$$

$$-\frac{\hbar^2}{2m}\frac{\partial^2 \psi_{non}}{\partial x^2} = -\frac{\hbar^2}{2m}\left[\frac{\partial^2 \psi_{non}}{\partial \phi_{x,t}^2}\left(\frac{\partial \phi_{x,t}}{\partial x}\right)^2 + \frac{\partial \psi_{non}}{\partial \phi_{x,t}}\frac{\partial^2 \phi_{x,t}}{\partial x^2}\right] + Q_1 \tag{21}$$

where $\psi_{non}$ – the non-linear function, $\hat{P}_{non}$ – the non-linear operator of the momentum, $\hat{V}$ – the operator of potential energy, $Q_1 = \frac{i\hbar}{2m}\frac{\partial \psi_{non}}{\partial \phi_{x,t}}\frac{\partial S^2(t,x)}{\partial x^2}\phi_{x,t}$ – the complex quantum potential, $S(t,x)$ – the action.

The time evolution equation for $\psi_{non}$ can be expected to be valid only in the non-relativistic regime. There is a non-linear theory underlying quantum mechanics, when the non-linearity is important for the state and the complex quantum potential. This is a new differential equation.

The NPDE can be solved by the method of separation of "variables". There exists a set of solutions of the form $\psi_{non}[t,x,\phi_{x,t}(t,x)] = \psi_{non,t}[t,\phi_t(t)]\psi_{non,x}[x,\phi_x(x)] \equiv \psi_{non,t}\psi_{non,x}$. We introduce a separation constant $E_{non,}$ and write two equations

$$i\hbar \frac{1}{\psi_{non,t}} \frac{\partial \psi_{non,t}}{\partial \phi_t} \frac{\partial \phi_t}{\partial t} = E_{non} \qquad (22)$$

$$\frac{p_x^2}{2m\psi_{non,x}} \left( \frac{\partial^2 \psi_{non,x}}{\partial \phi_x^2} \phi_x^2 + \frac{\partial \psi_{non,x}}{\partial \phi_x} \phi_x \right) + \frac{i\hbar}{2m\psi_{non,x}} \frac{\partial \psi_{non,x}}{\partial \phi_x} \frac{\partial^2 S}{\partial x^2} \phi_x + V = E_{non} \qquad (23)$$

where $V$ – the time-independent potential energy, $E_{non}$ – the non-linear energy of particle with mass $m$. Next, the detaching of $d\psi_{non,t}/d\phi_t$ can be treated as a fraction and split into two differentials, $d\psi_{non,t}$ and $d\phi_t$. When $\psi_{non,t}$ is nonlinear the analysis of the solution of the equation (22) is sometimes helped by considering a linear quantum mechanics whose the eigenvalue equation is given by the equation (13). According to the equation (13), the equation (22) can be rewritten in the form

$$\frac{d\psi_{non,t}}{\psi_{non,t}} = \beta_{non} \frac{d\phi_t}{\phi_t} \qquad (24)$$

where $\beta_{non} = E_{non}/E$, $\beta_{non}$ – the parameter describing the non-linearity. The general solution of equation (24) is

$$\Psi_{non,t} = C_t \cdot \exp(\beta_{non} \ln \phi_t) = C_t \cdot \exp(-i\beta_{non} \omega \cdot t) \qquad (25)$$

where $C_t$ - the constant, $\omega$ – the angular frequency. Thus, if the value of $\beta_{non}$ differs from 1 then the equation predicts non-linear quantum phenomena. The time evolution of the non-linear state is also given by the fifth postulate.

**Postulate 5**. *The time evolution of a non-linear quantum system preserves the normalisation of the associated vector ket. The time evolution of the non-linear function of a non-linear quantum system is described by* $|\psi_{non}[t,\phi(t)]> = \hat{U}(t,t_0)|\psi_{non}[t_0,\phi(t_o)]>$, *for some unitary operator* $\hat{U}$.

The unitary operator $\hat{U}_t := \exp(-iE_{non} \cdot \omega \cdot t/E)$ is easy to understand: after defining $\psi_{non}[t,\phi(t)] = \exp(-iE_{non} \cdot t/\hbar)\psi_{non}[t_0,\phi(t_o))$, one would compute as follows:

$$i\hbar \frac{\partial \psi_{non}[t,\phi(t)]}{\partial t} = i\hbar \frac{\partial \{\exp(-iE_{non} \cdot t/\hbar)\psi_{non}[t_0,\phi(t_o)]\}}{\partial t} =$$
$$= E_{non}\{\exp(-iE_{non} \cdot t/\hbar)\psi_{non}[t_0,\phi(t_o)]\} = E_{non}\psi_{non}[t,\phi(t)] \qquad (26)$$

This calculation can actually be justified for all $\psi_{non}[t,\phi(t)]$ in the domain of $E_{non}$.

**Proposition 5.1**. *(Implicit wave function in linear quantum mechanics). Let us consider an implicit wave function of* $\phi_{x,t} = \phi(t,x) = \phi_t(t)\phi_x(x)$ *and assume the existence of a constraint*

$$\psi_{non,t}[t,\phi_t(t)] = const \quad \text{and} \quad \psi_{non,x}[x,\phi_x(x)] = const \qquad (27)$$

which relates the $\phi_t(t)$ and $\phi_x(x)$ to the independent variables $t$ and $x$. Since $t$ and $x$ are independent variables, total differential must vanish ($d\,\psi_{non,t}[t,\phi_t(t)]=0$ and $d\psi_{non,x}[x,\phi_x(x)]=0$) and the derivatives of $\phi_t(t)$ and $\phi_x(x)$ are given by

$$\frac{\partial \phi_t(t)}{\partial t} = -\frac{\left(\frac{\partial \psi_{non,t}}{\partial t}\right)_{\phi_t}}{\left(\frac{\partial \psi_{non}}{\partial \phi_t}\right)_t} \quad and \quad \frac{\partial \phi_x(x)}{\partial x} = -\frac{\left(\frac{\partial \psi_{non}}{\partial x}\right)_{\phi_x}}{\left(\frac{\partial \psi_{non}}{\partial \phi_x}\right)_x} \tag{28}$$

where $\phi_t = \phi_t(t)$ – the time-dependent wave function, $\phi_x = \phi_x(x)$ – the position-dependent wave function.

The implicit function theorem gives sufficient conditions on the non-linear function $\psi_{non,x}[x,\phi(x)]$ ($\psi_{non,t}[t,\phi(t)]$) so that the equation can be solved for $\phi(x)$ ($\phi(t)$). The above formula comes from using the generalised chain rule to obtain the total derivative with respect to $t$ ($x$). For the separated wave, the function is well known to be correct, but for the exact wave the function is calculated by $\phi_{x,t}(t,x) = \phi_t(t)\phi_x(x)$. The results are a necessary and sufficient condition for the correspondence principle between the non-linear wave function and the wave function.

**Proposition 5.2**. *The non-linear quantum mechanics and linear quantum mechanics are equivalent when a free particle of mass m is described by the NPDE.*

For simplicity, we consider a free particle moving in a one-dimensional case when the potential energy vanishes. In this case, the NPDE becomes:

$$i\hbar\frac{\partial \psi_{non}}{\partial t} = -\frac{\hbar^2}{2m}\frac{\partial^2 \psi_{non}}{\partial x^2} \tag{29}$$

Thus, the task is to find out information about the implicit wave function from equation (24). The non-linear wave function solution can be put in the form

$$\psi_{non} = const \cdot \exp\{\frac{1}{\sqrt{2\pi\hbar}}\exp[\frac{i}{\hbar}(p_x x - Et)]\} \tag{30}$$

where $p_x$ – the eigenvalue of momentum operator, $E$ – the eigenvalue of energy operator. According to proposition 5.1, the implicit wave function solution is given by

$$\phi(t,x) = \frac{1}{\sqrt{2\pi\hbar}}\exp[\frac{i}{\hbar}(p_x x - Et)] + const \tag{31}$$

with eigenenergy equation

$$E = \frac{p_x^2}{2m} \tag{32}$$

The obtained results lead to exactly the same form for the linear quantum mechanics. In simple examples, it may be possible to give an explicit solution of NPDE, but usually that is a difficult problem.

The above argument is key and it is worth pointing out that it just uses the fact that a commutator plays a fundamental role in the physical interpretation of non-linear quantum mechanics. The last postulate is:

**Postulate 6.** *The commutator of position operators $\hat{x}^n$ and non-linear momentum operator $\hat{P}_{x,non}$ do not commute. Hence, the operators would satisfy*

$$[\hat{x}^n, \hat{P}_{x,non}]\psi_{non} = i\hbar n x^{n-1}\psi_{non} \qquad \text{for} \quad n = 1,2,3,... \tag{33}$$

where $[\hat{x}^n, \hat{P}_{x,non}] = \hat{x}^n \hat{P}_{x,non} - \hat{P}_{x,non} \hat{x}^n$ is the commutator of $\hat{x}^n$ and $\hat{P}_{x,non}$.

It is very important, however, to note that operator multiplication is not commutative. The commutator describes the consequences of sequences of measurements performed on a quantum system.

**Conclusion**

Introducing the quantum operators acting on the non-linear function in the non-relativistic regime, we propose the foundations of non-linear quantum mechanics based on six postulates and five propositions. Note that the implicit wave function can be determined by the non-linear function solution of NPDE. For this reason, we obtain a correspondence principle between the non-linear wave function and the wave function. The last result is a necessary and sufficient condition for the verified non-linear quantum mechanics. It should also be pointed out that the free particle problem and the fundamental commutation relation give good results known in science. The continuous non-linear quantum formalism can be formally generalised to the three-dimensional case and can be also rewritten in discrete notation.